\documentclass{birkmult}
\usepackage{amsmath}
\usepackage{graphicx}
\usepackage{amsfonts}
\usepackage{amssymb}

\theoremstyle{definition}

\theoremstyle{remark}

\numberwithin{equation}{section}

\begin{document}

\title{Variational Formulation for Quaternionic Quantum Mechanics}
\author{C. A. M. de Melo}
\address{%
Instituto de F\'{\i}sica Te\'{o}rica, UNESP - S\~{a}o Paulo State University.\\
Rua Pamplona 145, CEP 01405-900, S\~{a}o Paulo, SP, Brazil.\\ and}
\address{%
Universidade Vale do Rio Verde de Tr\^{e}s Cora\c{c}\~{o}es,\\
Av. Castelo Branco, 82 - Ch\'{a}cara das Rosas, P.O. Box 3050,\\
CEP 37410-000, Tr\^{e}s Cora\c{c}\~{o}es, MG, Brazil.}

\email{cassius.anderson@gmail.com}

\author{B. M. Pimentel}
\address{%
Instituto de F\'{\i}sica Te\'{o}rica, UNESP - S\~{a}o Paulo State University.\\
Rua Pamplona 145, CEP 01405-900, S\~{a}o Paulo, SP, Brazil.}
\email{b.m.pimentel@gmail.com}

\date{2008}
\dedicatory{In honour of the 70th birthday of Prof. J. A. C. Alcar\'{a}s}

\begin{abstract}
A quaternionic version of Quantum Mechanics is constructed using the
Schwinger\'s formulation based on measurements and a Variational Principle.
Commutation relations and evolution equations are provided, and the results
are compared with other formulations.
\end{abstract}

\maketitle

%
%
%
%
%
%
%
%
%

\address{Instituto de F\'{\i}sica Te\'{o}rica, UNESP - S\~{a}o Paulo State University.\\
Rua Pamplona 145, CEP 01405-900, S\~{a}o Paulo, SP, Brazil.\\ and}
\address{Universidade Vale do Rio Verde de Tr\^{e}s Cora\c{c}\~{o}es,\\
Av. Castelo Branco, 82 - Ch\'{a}cara das Rosas, P.O. Box 3050,\\
CEP 37410-000, Tr\^{e}s Cora\c{c}\~{o}es, MG, Brazil.}

\address{Instituto de F\'{\i}sica Te\'{o}rica, UNESP - S\~{a}o Paulo State University.\\
Rua Pamplona 145, CEP 01405-900, S\~{a}o Paulo, SP, Brazil.}


\section{Introduction\label{Quaternions}}

In 1936 Birkhoff and von Neumann \cite{Birkhoff-Neumann} have shown the
existence of a propositional calculus as fundamental ingredient of Quantum
Mechanics (QM), which could be written using only the outputs of measures.
It does not assume any set of numbers or even a particular vectorial space,
but contains the essentials of QM such as uncertainty relations and
complementary properties. Of course, the authors showed that there are three
different realizations for this propositional calculus, corresponding to the
real or complex numbers or still quaternions. Octonions and higher
dimensional extensions of the complex numbers are discarded, since they can
not have a conservation law for the probability current \cite{Adler}.

We can ask: which of these three realizations of the \textquotedblleft
general\textquotedblright\ QM of Birkhoff and von Neumann is present in
Nature? Here it is implicit the hypothesis that the set of numbers of a
given theory reflects part of the physical information about the system.
While the differences between the real and complex QM are relatively simple
and well known \cite{Stuckelberg}, the quaternionic version has many new and
rich characteristics. Therefore, it sounds strange that such possibility is
not much\ explored, but there are very good reasons for this. First, the
problem of writing a quaternionic Schr\"{o}dinger equation is not trivial
since it involves the explicit use of imaginary unit. Second, the
representation of composite systems by a direct product is more difficult
due to the noncommutativity of the quaternionic valued wave functions.

Here, we implement a quaternionic version of Schwinger's Measurement Algebra
and build the dynamics based on the Action Principle. In each step, the
analogy with the usual QM is used as inspiration, but the peculiarities
emerging from the quaternionic noncommutativity are always emphasized.

The theory constructed by this means is quite distinct from Adler's approach
\cite{Adler}, having similarities with the work of Finkelstein, Jauch,
Schiminovich and Speiser, \cite{FinkQ1, FinkQ2, FinkQ3}.

\section{Measurement Symbols}

The classical theory of physical measurements is based on the concept that
the interaction between the system under observation and the measurement
apparatus can be done arbitrarily small or, at least, precisely compensated,
in such way to specify an idealized measurement which does not disturb any
other property of the system. However, the experiment had demonstrated that
the interaction can not be done arbitrarily small neither the disturb
produced can be precisely compensated since it is uncontrollable and
unpredictable. The fact that the interaction can not be arbitrarily small is
expressed by the finite size of the Planck constant, while the
uncontrollable character of the interaction is given by the uncertainty
principle. Therefore, the measurement of a given property can produce a
significant change in the value of another previously measured property, and
then there is no sense in speaking about an microscopic system with definite
values for all its attributes. This is in contradiction with the classical
representation of physical quantities by numbers. The laws of a microscopic
physical system must then be expressed in a non-classical mathematical
language constituting a symbolic expression of the properties of microscopic
measurements.

In what follows, we will develop the general lines of such mathematical
structure discussing about simplified physical systems where any physical
quantity $A$ can have only a finite number of different values $%
a^{1},a^{2},a^{3}$ ... . The most simple measurement consider an ensemble of
similar independent systems which is divided by the apparatus of measurement
in sub-ensembles distinguished by the defined values of the physical
quantity under measurement. Let us denote $\hat{M}_{a}$ the selective
measurement accepting any system having value $a$ for the property $A$ and
rejecting any other. The addition of such symbols is defined as implying a
less specific measure, resulting in a sub-ensemble associated with any value
under the sum, none of them being distinguished of the others by the
measurement.

The multiplication of measurement symbols implies the sequence of
measurements reading from right to left. From the physical meaning of such
operations, we learn that addition is commutative and associative while
multiplication is only associative. Using $\hat{1}$ and $\hat{0}$ to
represent respectively the measures which accept and reject all systems, the
properties of the elementary selective measurement are given by\footnote{%
Of course, such properties characterize the measurement symbols as \emph{%
projectors} on the space of physical states. The projective geometry
originated from this complete set of projectors can be explored to construct
a pair of dual vector spaces of creation and anihilation operators
representing the \emph{out} and \emph{in} stages of an elementary
measurement.}
\begin{subequations}
\begin{align}
\hat{M}_{a}\hat{M}_{a} & =\hat{M}_{a}  \label{medida1} \\
\hat{M}_{a}\hat{M}_{a%
{\acute{}}%
} & =\hat{0}  \label{medida2} \\
\sum_{a}\hat{M}_{a} & =\hat{1}   \label{medida3}
\end{align}

From the meaning of the measurements represented by $\hat{1}$ and $\hat{0}$
we directly read the following algebraic properties:
\end{subequations}
\begin{align*}
\hat{1}\hat{M}_{a} & =\hat{M}_{a}\hat{1}=\hat{M}_{a} \\
\hat{0}\hat{M}_{a} & =\hat{M}_{a}\hat{0}=\hat{0} \\
\hat{M}_{a}+\hat{0} & =\hat{M}_{a}
\end{align*}
what justifies the adopted notation. The algebraic properties of $\hat{1}$, $%
\hat{0}$ and $\hat{M}_{a}$ are consistent provided that the multiplication
be distributive,
\begin{equation*}
\sum_{a}\left( \hat{M}_{a}\hat{M}_{a%
{\acute{}}%
}\right) =\hat{M}_{a%
{\acute{}}%
}=\hat{M}_{a%
{\acute{}}%
}\hat{1}=\hat{M}_{a%
{\acute{}}%
}\sum_{a}\hat{M}_{a}
\end{equation*}

All laws of multiplication for measurement symbols given above can be
combined in a single expression,
\begin{equation*}
\hat{M}_{a}\hat{M}_{a%
{\acute{}}%
}=\delta_{\,a%
{\acute{}}%
}^{a}\hat{M}_{a}
\end{equation*}
with the introduction of the symbol
\begin{equation*}
\delta_{\,a%
{\acute{}}%
}^{a}=\left\{
\begin{array}{c}
\hat{1},\qquad a=a%
{\acute{}}
\\
\hat{0},\qquad a\not =a%
{\acute{}}%
\end{array}
\right.
\end{equation*}
known as\emph{\ Kronecker's} \emph{delta}.

From these definitions one sees that the measurement symbols belong to a
noncommutative ring \cite{Anel}.

\section{Compatible Properties}

Two quantities $A_{1}$ and $A_{2}$ are compatible when the measurement of
one of them does not destroy the knowledgement of a previous measurement of
the other. The selective measures $\hat{M}_{a_{1}}$ and $\hat{M}_{a_{2}}$,
taken in this order, produce an ensemble where it is possible, simultaneously%
\footnote{%
Note that the use of the word \emph{simultaneously} is made without any
reference to a definition of \emph{simultaneity} and also without reference
to the concept of \emph{time}. Here, we are presuming that in an intuitive
way it is clear to the reader the sense in which these words are been used.
The concept of temporal evolution is associated with the notion of dynamics
which will be investigated below based on the Action Principle.}, to
attribute the values $a_{1}$ to $A_{1}$ and $a_{2}$ to $A_{2}$. The symbol
for such composite measurement is
\begin{equation*}
\hat{M}_{a_{1}a_{2}}=\hat{M}_{a_{1}}\hat{M}_{a_{2}}=\hat{M}_{a_{2}}\hat {M}%
_{a_{1}}
\end{equation*}
From such definition it is easy to see that the compatibility is an
equivalence relation.

A complete set $A$ of compatible quantities $A_{1},...,A_{r}$ means that any
pair of such properties is compatible and there is no other compatible
quantity outside the set, except the functions constructed from the set $A$.
In fact, $A$ is an \emph{equivalence class}. The measurement symbol
\begin{equation*}
\hat{M}_{a}=\prod_{r}\hat{M}_{a_{r}}
\end{equation*}
describes a complete measurement where the selected systems have definite
values for the maximum number of possible attributes. Any tentative for
determining the value of another independent physical quantity will produce
uncontrollable changes on the previously measured values. Therefore, the
optimum information about a given system is achieved making a complete
selective measurement. The systems accepted by the complete selective
measurement $\hat{M}_{a}$ are known to being in the state $a$. The symbolic
properties for the complete measures are the same as for the elementary
selective measurements, i.e., (\ref{medida1}), (\ref{medida2}) and (\ref%
{medida3}).

\section{Changing States Measurements}

A more general kind of measure incorporates a change on the state of the
system. The symbol $\hat{M}_{a}^{a_{1}}$ represents a complete selective
measurement which accepts systems in the $a_{1}$ state and let out systems
in the state $a$. The measurement process $\hat{M}_{a}$ is the special case
when no change on the state occurs,%
\begin{equation*}
\hat{M}_{a}=\hat{M}_{a}^{a}
\end{equation*}

The properties of successive measurements of this specie are given by
\begin{equation}
\hat{M}_{a_{1}}^{a_{2}}\hat{M}_{a_{3}}^{a_{4}}=\delta_{%
\,a_{3}}^{a_{2}}M_{a_{1}}^{a_{4}}   \label{medidaprod}
\end{equation}
since if $a_{3}\not =a_{2}$ the second stage of the apparatus does not
select any system emerging from the first one, and if $a_{3}=a_{2}$ all
systems coming from the first stage are accepted by the second, being the
composite measurement a selection of systems in the state $a_{4}$ and
letting it out in the state $a_{1}$. Observe that if we interchange both
stages, then
\begin{equation*}
\hat{M}_{a_{3}}^{a_{4}}\hat{M}_{a_{1}}^{a_{2}}=\delta_{%
\,a_{1}}^{a_{4}}M_{a_{3}}^{a_{2}}
\end{equation*}
what is not the same as (\ref{medidaprod}). Therefore, we realize that the
multiplication of complete measurements symbols is noncommutative.

The physical quantities belonging to a complete set do not exhaust the
totality of physical attributes in a system. One can form others complete
sets $B,C,$ ..., which are mutually incompatible and, for each choice of
non-interfering physical characteristics, there is a set of selective
measurements concerning to systems in the appropriate states $\hat{M}%
_{b_{1}}^{b_{2}},\ \hat{M}_{c_{1}}^{c_{2}},$ ... The most general selective
measurement links two complete sets of incompatible properties. Let $\hat {M}%
_{a}^{b}$ be the measurement process rejecting all systems which are not in
the state $b$ and allowing to emerge only systems in the state $a$. The
composite measurement $\hat{M}_{a}^{b}\hat{M}_{c}^{d}$ will select systems
in the state $d$ and let them in the state $a$, so it should be proportional
to the selective measurement $\hat{M}_{a}^{d}$.

The examples considered until now include the passing of all or none system
through both stages, as realized by the symbols $\hat{1}$ and $\hat{0}$.
Notwithstanding, in general we can just admit that measures of the property $%
B$ upon a system in the state $c$, which belongs to a complete set
incompatible with $B$, will furnishes an statistical distribution of all
possible results. So, only a fraction of the systems emerging from the first
stage is accepted by the second one. We can express this by the general
multiplication law:
\begin{equation}
\hat{M}_{a}^{b}\hat{M}_{c}^{d}=\left\vert a\right\rangle \left\langle
b|c\right\rangle \left\langle d\right\vert =\hat{M}_{a}^{d}\left(
\left\langle b|c\right\rangle \right)   \label{prodgeral}
\end{equation}
where $\left\langle b|c\right\rangle $ is a number characterizing the
statistical relationship between the states $b$ and $c$. In particular,
\begin{equation*}
\left\langle a|a%
{\acute{}}%
\right\rangle =\delta_{\,a%
{\acute{}}%
}^{a}\qquad a,a%
{\acute{}}%
\sqsubset A
\end{equation*}
where $\sqsubset$ means that $a$ and $a%
{\acute{}}%
$ are defined sets of values for the complete set $A$. Since that the
numbers $\left\langle a|b\right\rangle $ link the states $a$ and $b$ they
are called \emph{transformation function}.

The measurement symbols $M_{a}^{b}$ equipped with addition and
multiplication as defined above and together with the scalar ring $%
\left\langle b|c\right\rangle $ form an algebra, which we call the \emph{%
Measurement Algebra}. Observe that nothing was said about the particular set
of numbers $\left\langle b|c\right\rangle $\ to be adopted. In fact, as
matter for mathematical and physical meaning consistency, it is enough that $%
\left\langle b|c\right\rangle $ belongs to an scalar ring.

Of course, the order in which the scalars $\left\langle a|b\right\rangle $\
appear in the product (\ref{prodgeral}) is very important, since it reflects
on the ring multiplication law, allowing the definition of different
measurement algebras. Therefore, the most general form to indicate the
multiplication rule for measurement symbols is $\hat{M}_{a}^{b}\hat{M}%
_{c}^{d}=\hat{M}_{a}^{d}\left( \left\langle b|c\right\rangle \right) $ since
it does not make any reference to the order of the scalar on the product.
However, we will maintain the scalars on a preferable central position on
the product. Our main interest here is to suppose that the scalars are
quaternions and investigate what are the physical implications of such
assumption.

The reason to take the scalars on a central multiplicative position comes
from the recognition that measurement symbols are in fact projectors on the
several possible states of two different complete sets of observables. To
reinforce such character, we adopt the notation%
\begin{equation*}
\hat{M}_{a}^{b}=\left\vert a\right\rangle \left\langle b\right\vert
\end{equation*}

Then, the most general way in which a measurement symbol can appear together
an scalar is%
\begin{equation*}
\hat{M}_{a}^{b}\left( q\right) =\left\vert a\right\rangle q\left\langle
b\right\vert
\end{equation*}
being $q$ any element of the ring under which the measurement algebra is
defined. As stated before, we will assume that the numbers $q$ are \emph{%
quaternions}, defined by
\begin{equation*}
q=q_{0}+q_{1}e_{1}+q_{2}e_{2}+q_{3}e_{3}\,,\,e_{i}e_{j}=-\delta_{ij}+%
\sum_{k=1}^{3}\varepsilon_{ijk}e_{k}\,,\,q_{n}\in\mathbb{R~}\forall
~n\in\left\{ 0,...,3\right\}
\end{equation*}
When $q=1$ we simply denote $\hat{M}_{a}^{b}\left( 1\right) =\hat{M}_{a}^{b}$%
. This notation is useful because it maintains separated in an explicitly
way the two parts of the measurement symbol corresponding to the physical
Hilbert space of states $\mathcal{H}$ and its dual $\mathcal{H}^{\dagger}$.
In the language of second quantization, this notation directly alludes to
the annihilation (right) and creation (left) processes of particles or field
fluctuations involved in a measurement act. It is important to stand out
that since the products of vector \emph{by} scalars are defined over a
noncommutative ring, these products have sense only a \emph{definite} order,
which we take as right for the kets $\left( \left\vert a\right\rangle
q,\,\forall\left\vert a\right\rangle \in\mathcal{H}\left( \mathbb{H}\right)
,\,\forall q\in\mathbb{H}\right) $ and left for the bras $\left(
q\left\langle b\right\vert ,\,\forall\left\langle b\right\vert \in \mathcal{H%
}^{\dagger}\left( \mathbb{H}\right) ,\,\forall q\in\mathbb{H}\right) $,
where $\mathcal{H}\left( \mathbb{H}\right) $ is the Hilbert space of
eigenstates of a given complete set of observables.

Quaternions are a particular realization of a \emph{Clifford algebra} \cite%
{Lounesto}, so a even more general theory can be recognized.

\section{Transformation Functions}

The fundamental transformation law for the measurement symbols is
essentially unaffected by the specific choice of the scalar ring. Actually,
using the notation of the previous section, measurement symbols of one kind
can be transformed in symbols of another kind:
\begin{equation}
\hat{M}_{c}^{d}=\left\vert c\right\rangle \left\langle d\right\vert
=\sum_{a,b}\hat{M}_{a}\hat{M}_{c}^{d}\hat{M}_{b}=\sum_{a,b}\left\vert
a\right\rangle \left\langle a|c\right\rangle \left\langle d|b\right\rangle
\left\langle b\right\vert   \label{TransfSimbMedQ}
\end{equation}

Carefully preserving the composition of products, one can interpret this
relation as a double mapping of vectors $\left\vert c\right\rangle $\ and
covectors\ $\left\langle d\right\vert $ on the linear combinations $\sum
_{a}\left\vert a\right\rangle \left\langle a|c\right\rangle $\ and $\sum
_{b}\left\langle d|b\right\rangle \left\langle b\right\vert $ respectively.
Therefore, the composition law for transformation functions in a
quaternionic ring is
\begin{equation*}
\sum_{b}\left\langle a|b\right\rangle \left\langle b|c\right\rangle
=\left\langle a|c\right\rangle
\end{equation*}
from which we obtain the completeness relations%
\begin{align*}
\sum_{a}^{N}\sum_{b}^{N%
{\acute{}}%
{\acute{}}%
}\left\langle a|b\right\rangle \left\langle b|a\right\rangle &
=\sum_{a}^{N}1=N \\
\sum_{b}^{N%
{\acute{}}%
{\acute{}}%
}\sum_{a}^{N}\left\langle b|a\right\rangle \left\langle a|b\right\rangle &
=\sum_{b}^{N%
{\acute{}}%
{\acute{}}%
}1=N%
{\acute{}}%
\end{align*}

However, since quaternions do not commute, the preservation of the number of
degrees of freedom imply that%
\begin{equation}
\sum_{a}^{N%
{\acute{}}%
{\acute{}}%
}\sum_{b}^{N}\left\langle a|b\right\rangle \left\langle b|a\right\rangle
=\sum_{b}^{N}\sum_{a}^{N%
{\acute{}}%
{\acute{}}%
}\left\langle b|a\right\rangle \left\langle a|b\right\rangle   \label{comutQ}
\end{equation}
Except for systems with only one degree of freedom, this does not mean that $%
\left\langle a|b\right\rangle \left\langle b|a\right\rangle =\left\langle
b|a\right\rangle \left\langle a|b\right\rangle $ for any pair of
quaternionic transformation functions. Then, the relation (\ref{comutQ})
implies a restriction, but its interpretations is not easy.

\section{The Trace Functional and the Statistical Interpretation\label%
{estatistica}}

One of the most important actions over the measurement algebra is the \emph{%
trace functional}, which associates each element of the algebra to one
scalar. Since here the scalar ring is noncommutative, there are three kinds
of trace functional called respectively \emph{left, right }and \emph{central}
trace:%
\begin{equation*}
Tr_{L}\hat{M}_{a}^{b}\left( q\right) \equiv q\left\langle b|a\right\rangle
\end{equation*}%
\begin{equation*}
Tr_{R}\hat{M}_{a}^{b}\left( q\right) \equiv\left\langle b|a\right\rangle q
\end{equation*}%
\begin{equation*}
Tr_{C}\hat{M}_{a}^{b}\left( q\right) \equiv\sum_{\in}\left\langle
e\right\vert \left\vert a\right\rangle q\left\langle b\right\vert \left\vert
e\right\rangle
\end{equation*}

In the standard complex case, the trace functional is related to the
statistical interpretation of quantum mechanics. Here we have a more
complicated situation since none of the above trace functionals has an
invariant law of transformation. Nevertheless, the multiplication law is
invariant under the following mapping:
\begin{subequations}
\label{TransfSemeQ}
\begin{align}
\hat{M}_{a}^{b} & =\left\vert a\right\rangle \left\langle b\right\vert
\rightarrow\left\vert a\right\rangle \lambda_{a}^{-1}\lambda_{b}\left\langle
b\right\vert =\hat{M}_{a}^{b}\left( \lambda_{a}^{-1}\lambda_{b}\right)
\label{TransfSeme1Q} \\
\left\langle a|b\right\rangle & \rightarrow\lambda_{a}\left\langle
a|b\right\rangle \lambda_{b}^{-1}   \label{TransfSeme2Q}
\end{align}
where quaternions $\lambda_{a}$, $\lambda_{b}$ are not null. Therefore, the
transformation function $\left\langle a|b\right\rangle $ can not itself have
a direct physical interpretation, and shall configure in a combination
invariant under (\ref{TransfSemeQ}).

The appropriate basement for the statistical interpretation of the
transformation function can be inferred from a sequence of elementary
measurement, $\hat{M}_{b}\hat{M}_{a}\hat{M}_{b}$, which differs from $\hat {M%
}_{b}$ only by virtue of the disturbance caused by the intermediary
measurement of the attribute $A$. Only a fraction of the systems selected by
the initial measurement of $B$ is transmitted through the complete set.
Hence, we obtain the following symbolic statement:
\end{subequations}
\begin{equation*}
\hat{M}_{b}\hat{M}_{a}\hat{M}_{b}=\hat{M}_{b}\left( p\left( a|b\right)
\right)
\end{equation*}
where the number
\begin{equation}
p\left( a|b\right) =\left\langle b|a\right\rangle \left\langle
a|b\right\rangle   \label{ProbTrans}
\end{equation}
should be invariant under (\ref{TransfSemeQ}). It means that%
\begin{equation*}
\lambda_{b}\left\langle b|a\right\rangle \left\langle a|b\right\rangle
\lambda_{b}^{-1}=\left\langle b|a\right\rangle \left\langle a|b\right\rangle
\end{equation*}

Now, if one considers a measurement of the property $A$ which does not
distinguish between two states, one arrives on the additivity of $p\left(
a|b\right) ,$%
\begin{equation*}
\hat{M}_{b}\left( \hat{M}_{a}+\hat{M}_{a^{\prime }}\right) \hat{M}%
_{b}=\left( p\left( a|b\right) +p\left( a^{\prime }|b\right) \right) \hat{M}%
_{b}
\end{equation*}%
So, taking a measurement of $A$ unable to select any of such states, one
obtains
\begin{equation*}
\hat{M}_{b}\left( \sum_{a}\hat{M}_{a}\right) \hat{M}_{b}=\hat{M}_{b}
\end{equation*}%
what implies:
\begin{equation*}
\sum_{a}p\left( a|b\right) =1
\end{equation*}%
Such properties characterize $p\left( a|b\right) $ as a \emph{probability
measure} \cite{Probability} of observing the state $a$ in a measurement made
over a system known to be in the state $b$. However, probability measures
are positive real numbers, then we must to impose a restriction on the the
numbers which figure in the measurement algebra. Until now, all we have made
can be applied equally to quaternions or complex numbers. In fact, no
physical information was used to select the nature of such numbers, being
only necessary they form a scalar ring in order to obtain an algebra from
the elementary selective measurements. Therefore, any field, as $\mathbb{R}$
or $\mathbb{C}$, for instance, is candidate to figure as scalars in this
construction of the quantum theory, but also a ring which is not a field, as
quaternions or octonions, could be used. The extension of Quantum Mechanics
that we want to do here is to get quaternions as the scalar ring used to
construct the measurement algebra.

So, the probability measure $p\left( a|b\right) $ must satisfy $p\left(
a|b\right) \geqslant0$. Besides, the arbitrary reading convention in the
multiplicative law implies that such probability shall be \emph{symmetric}.
The simplest way to accomplish all these properties is to demand $Q=\lambda
_{b}\left\langle b|a\right\rangle $ and $\bar{Q}=\left\langle
a|b\right\rangle \lambda_{b}^{-1}$ to be a conjugated pair. Of course, in
such case one obtains%
\begin{equation*}
Q\bar{Q}=\bar{Q}Q=\left\vert Q\right\vert ^{2}\geqslant0
\end{equation*}%
\begin{equation*}
\lambda_{b}\left\langle b|a\right\rangle \left\langle a|b\right\rangle
\lambda_{b}^{-1}=\left\langle a|b\right\rangle \lambda_{b}^{-1}\lambda
_{b}\left\langle b|a\right\rangle =\left\langle a|b\right\rangle
\left\langle b|a\right\rangle
\end{equation*}

On the other hand,%
\begin{equation*}
\bar{Q}=\overline{\left( \lambda_{b}\left\langle b|a\right\rangle \right) }=%
\overline{\left\langle b|a\right\rangle }\bar{\lambda}_{b}=\left\langle
a|b\right\rangle \lambda_{b}^{-1}
\end{equation*}
This let us with the following statements:%
\begin{align*}
\left\langle b|a\right\rangle \left\langle a|b\right\rangle & =\left\langle
a|b\right\rangle \left\langle b|a\right\rangle \geqslant0 \\
\overline{\left\langle b|a\right\rangle }\bar{\lambda}_{b} & =\left\langle
a|b\right\rangle \lambda_{b}^{-1}
\end{align*}

Again, the simplest way to solve this system is taking%
\begin{align*}
\bar{\lambda}_{b} & =\lambda_{b}^{-1} \\
\overline{\left\langle b|a\right\rangle } & =\left\langle a|b\right\rangle
\end{align*}
With this choice one is able to recover all the properties of the
probability measure $p\left( a|b\right) $.

Using an exponential representation for $\lambda_{a}$ we see that the first
condition above can be written in the form
\begin{equation*}
\lambda_{a}=Ae^{e_{\lambda}\varphi\left( a\right) }\rightarrow
Ae^{-e_{\lambda}\varphi\left( a\right) }=A^{-1}e^{-e_{\lambda}\varphi\left(
a\right) }\rightarrow A^{2}=1\rightarrow A=\pm1
\end{equation*}
where
\begin{align*}
\left\vert A\right\vert & =\left\vert \lambda_{a}\right\vert =\left[ \left(
\lambda_{a}^{0}\right) ^{2}+\left( \lambda_{a}^{1}\right) ^{2}+\left(
\lambda_{a}^{2}\right) ^{2}+\left( \lambda_{a}^{3}\right) ^{2}\right] ^{1/2}
\\
e_{\lambda} & =\frac{\lambda_{a}^{1}e_{1}+\lambda_{a}^{2}e_{2}+\lambda
_{a}^{3}e_{3}}{\left[ \left( \lambda_{a}^{1}\right) ^{2}+\left(
\lambda_{a}^{2}\right) ^{2}+\left( \lambda_{a}^{3}\right) ^{2}\right] ^{1/2}}
\\
\varphi\left( a\right) & =\arctan\left( \frac{\lambda_{a}^{0}}{\left\vert
\lambda_{a}\right\vert }\right) \quad\varphi\left( a\right) \in\left[ 0,\pi%
\right]
\end{align*}
the choice for the signal in $A$ is arbitrary and no physical effect can be
distinguished by one particular choice. Therefore we will take the positive
signal. Since $\lambda_{a}$ is a unitary arbitrary number its phase $%
\varphi\left( a\right) $ can be an arbitrary real number.

Thus, besides the problems concerning about the definition of the trace
functional one is still able to construct an statistical interpretation for
the Quaternionic Quantum Mechanics. In fact, such result indicates that the
roots for the statistical interpretation are in the propositional calculus%
\footnote{%
Or, in our construction, in the Measurement Algebra relations.} of Birkhoff
and von Neumann \cite{Birkhoff-Neumann}, and not in the particular system of
numbers adopted to construct the theory.

Another very important piece for the construction of the statistical
interpretation was the automorphism $\left\langle a|b\right\rangle
\rightarrow\lambda_{a}\left\langle a|b\right\rangle \lambda_{b}^{-1}$\ for
the scalar ring $\mathbb{H}$. But, physically, what means such
identification? We know that the elements of the scalar ring represent
logical relations between the possible physical states of the system under
consideration. Clearly, it is even possible to say when two of such relation
are \textquotedblleft the same thing\textquotedblright\ for states taken in
distinct physical systems without departing the traditional concepts of pure
logic\footnote{%
The role for the abstract mathematical logic in Physics is discussed in a
very interesting way in \cite{Manin}.}, i.e., without using the concepts of
structured networks introduced by Birkhoff and von Neumann \cite%
{Birkhoff-Neumann}. However, this defines such numbers modulo automorphisms
\cite{FinkQ1}. In the case of a quantum theory with only real numbers this
is sufficient to determine completely such numbers \cite{Stuckelberg}. In
the complex case, it still stands an ambiguity, which is manifested under the
existence of a conjugated algebra. In Quaternionic Quantum Mechanics such
ambiguousness is infinitely bigger. This requires the introduction of more
structure elements in the theory\footnote{%
Of course, these observations are crucial to construct the representation
for systems with many particles.}. In the following we will delimitate what
are suchs structures.

\section{The Adjoint}

Other important aspect of the probabilistic interpretation for (\ref%
{ProbTrans}) is the symmetry
\begin{equation*}
p\left( a|b\right) =p\left( b|a\right)
\end{equation*}

Remember the arbitrary convention for reading the measurement symbols and
their products: the order of the events is read from right to left. But any
equation involving the measurement symbols is equally valid if interpreted
in the opposite sense and none physical result can depend of what is the
convention adopted. Introducing the right-handed interpretation, $%
\left\langle a|b\right\rangle $ acquire the same meaning of $\left\langle
b|a\right\rangle $ in the left-handed convention. We can conclude that the
probability connecting the states $a$ and $b$ in a given sequence must be
constructed symmetrically from $\left\langle a|b\right\rangle $ and $%
\left\langle b|a\right\rangle $. Of course, this is the reason why $p\left(
a|b\right) $\ should be symmetric. The introduction of the opposite
convention for the measurement symbols will be called the \emph{adjoint}
operation and will denoted by $^{\dagger}$. Therefore,
\begin{equation*}
\hat{M}_{a}^{b\dagger}=\hat{M}_{b}^{a}
\end{equation*}
and
\begin{equation*}
M_{a%
{\acute{}}%
}^{a\dagger}=M_{a}^{a%
{\acute{}}%
}
\end{equation*}

in particular,
\begin{equation*}
M_{a}^{\dagger}=M_{a}
\end{equation*}
what means that $\hat{M}_{a}$ is a self-adjoint operator. For the product of
measurements symbols we have
\begin{equation*}
\left( \hat{M}_{a}^{b}\hat{M}_{c}^{d}\right) ^{\dagger}=\hat{M}_{d}^{c}\hat{M%
}_{b}^{a}=\hat{M}_{c}^{d\dagger}\hat{M}_{a}^{b\dagger}
\end{equation*}

The meaning of addition is not changed by the adjoint operation what permits
to extend these properties for all element in the measurement algebra:
\begin{equation*}
\left( \hat{X}+\hat{Y}\right) ^{\dagger}=\hat{X}^{\dagger}+\hat{Y}^{\dagger
}\qquad\left( \hat{X}\hat{Y}\right) ^{\dagger}=\hat{Y}^{\dagger}\hat {X}%
^{\dagger}\qquad\left( \lambda\hat{Y}\right) ^{\dagger}=\hat{Y}^{\dagger }%
\bar{\lambda}
\end{equation*}
where $\lambda\in\mathbb{H}$.

\section{Infinitesimal Variation of Transformation Functions\label{Condicoes}%
}

Taking infinitesimal variations of the two fundamental properties of the
transformations functions, we find%
\begin{align}
\sum_{b}\left[ \delta\left\langle a|b\right\rangle \left( \left\langle
b|c\right\rangle \right) +\left\langle a|b\right\rangle \delta\left\langle
b|c\right\rangle \right] & =\delta\left\langle a|c\right\rangle
\label{DifFuncTrans2} \\
\delta\overline{\left\langle a|b\right\rangle } & =\delta\left\langle
b|a\right\rangle  \notag
\end{align}

In the ordinary complex case \cite{Schwinger} the numbers $%
\delta\left\langle a|b\right\rangle $\ are interpreted as representing the
matrix elements of an infinitesimal operator,%
\begin{equation*}
\delta\left\langle a|b\right\rangle =i\left\langle a\right\vert \delta\hat {W%
}_{ab}\left\vert b\right\rangle
\end{equation*}
where the constant $i$ was chosen in order to assure that the operator $%
\delta\hat{W}_{ab}$\ is self-adjoint.

Here, it is an open question what constant should be chosen since actually
we have \emph{tree} imaginary unities. The most general case is let the
imaginary unity to be an \emph{operator }$\hat{\iota}$ where $i\hat{1}=\hat{%
\iota}$ can be considered as a particular case for $\mathbb{C}$.

Let it be so, defining%
\begin{equation}
\delta\left\langle a|b\right\rangle =\left\langle a\right\vert \hat{\iota }%
\delta\hat{W}_{ab}\left\vert b\right\rangle   \label{OperInf2}
\end{equation}
where $\hat{\iota}$ is a quaternionic valued operator that we will be fixed
later under the requirement of $\delta\hat{W}_{ab}$\ be a self-adjoint
operator. Using this definition it is easy to see that the additivity and
the skewsymmetry in ordering infinitesimal operators are the same as in the
complex case \cite{Schwinger},%
\begin{equation*}
\delta\hat{W}_{ac}=\delta\hat{W}_{ab}+\delta\hat{W}_{bc}
\end{equation*}%
\begin{equation*}
\delta\hat{W}_{ba}=-\delta\hat{W}_{ab}
\end{equation*}

On the other hand,%
\begin{equation*}
\delta\overline{\left\langle a|b\right\rangle }=\left\langle b\right\vert
\delta\hat{W}_{ab}^{\dagger}\hat{\iota}^{\dagger}\left\vert a\right\rangle
=\left\langle b\right\vert \hat{\iota}\delta\hat{W}_{ba}\left\vert
a\right\rangle
\end{equation*}
what let us to the operatorial identity,%
\begin{equation*}
\delta\hat{W}_{ab}^{\dagger}\hat{\iota}^{\dagger}+\hat{\iota}\delta\hat {W}%
_{ab}=\hat{0}
\end{equation*}

If we impose%
\begin{equation}
\delta\hat{W}_{ab}=\delta\hat{W}_{ab}^{\dagger}   \label{AutoAdjQ}
\end{equation}%
\begin{equation}
\left[ \hat{\iota},\delta\hat{W}_{ab}\right] =\hat{0}   \label{Superselecao}
\end{equation}
we obtain:%
\begin{equation*}
\hat{\iota}=-\hat{\iota}^{\dagger}
\end{equation*}

This identity can be interpreted as a generalization of the complex
conjugation over $\mathbb{C}$, and shows that the operator $\hat{\iota}$\
behaves like an \textquotedblleft imaginary unit\textquotedblright. The
condition (\ref{AutoAdjQ}) assures the reality of the spectrum associated to
infinitesimal operators. The condition (\ref{Superselecao}) can be satisfied
in several ways:

\begin{enumerate}
\item demanding that all infinitesimal operator commutes with the imaginary
unity;

\item letting the imaginary unity to commute with any operator;

\item claiming that an infinitesimal operator commutes with any other
operator.
\end{enumerate}

In the standard quantum mechanics Schwinger choose the last option \cite%
{Schwinger}, which was subsequently extended to more general variations by
several authors \cite{Variacoes}. Here, we can see no reason to discard the
other two options. In fact, in their work on quaternionic quantum theory,
Finkelstein, Jauch Schiminovich and Speiser \cite{FinkQ1} have adopted a
particular case of the second option in the list above interpreting it as a
superselection rule\footnote{%
In \cite{FinkQ1} the imaginary unity operator is denoted by $\hat{\eta}$.}.
For while, we will require that at least one of the three conditions above
is satisfied, i.e., we will work directly assuming only the general
statement (\ref{Superselecao}).

With these choices, unitary infinitesimal operators can be expressed as%
\begin{equation*}
\hat{U}=\hat{1}+\hat{G},\qquad\hat{U}^{\dagger}=\hat{U}^{-1}=\hat{1}-\hat {G}%
,\qquad\hat{G}=-\hat{G}^{\dagger}=\hat{\iota}\delta\hat{W}
\end{equation*}
and infinitesimal variations of operators are induced by the commutator with
the generator%
\begin{equation}
\delta\hat{X}=-\left[ \hat{X},\hat{G}\right] =\left[ \hat{G},\hat {X}\right]
\label{ComVarInd2}
\end{equation}

These are all ingredients necessary to describe completely the one particle
physical states. We will not approach here the problem of representing
composite systems, but it is clear that such extension is possible. Now we
are ready to analyse the \emph{dynamic} characteristics which are changed by
the use of quaternions.

\section{The Variational Principle}

The quantum dynamics for the system will be obtained from the Schwinger
Action Principle \cite{Schwinger} here expressed as%
\begin{equation*}
\delta\left\langle a_{t_{2}}|b_{t_{1}}\right\rangle =\left\langle
a_{t_{2}}\right\vert \hat{\iota}\delta\hat{S}_{t_{1},t_{2}}\left\vert
b_{t_{1}}\right\rangle
\end{equation*}%
\begin{equation*}
\delta\hat{S}_{t_{1},t_{2}}=\left[ \mathbf{\hat{p}}\cdot\delta\mathbf{\hat {q%
}-}\hat{H}\delta t\right] _{t_{1}}^{t_{2}}+\int_{t_{1}}^{t_{2}}dt\frac {\vec{%
\delta}\hat{L}}{\vec{\delta}\mathbf{\hat{q}}}\cdot\left( \delta \mathbf{\hat{%
q}-}\widehat{\mathbf{\dot{q}}}\delta t\right) =\hat{G}_{2}-\hat{G}_{1}
\end{equation*}%
\begin{equation*}
\mathbf{\hat{p}}=\frac{\vec{\partial}\hat{L}}{\vec{\partial}\widehat {%
\mathbf{\dot{q}}}},\quad\hat{H}=\mathbf{\hat{p}}\cdot\widehat{\mathbf{\dot {q%
}}}-\hat{L}
\end{equation*}

The Hamiltonian $\hat{H}$\ and Lagrangian $\hat{L}$\ operators are
self-adjoints.

Schwinger Action Principle is the quantum counterpart of the classical Weiss
Principle \cite{Weiss}, which can be considered the most general variational
principle for classical fields. Schwinger Principle has been successfully
applied in Minkowiski \cite{QuanField}, curved \cite{DeWitt} or torsioned
spaces \cite{DKP}, as well as to describe quantum gauge transformations \cite%
{Bfield} and many other problems. Here, we will apply the Action Principle
to extract dinamic and kinematic information from a canonical formulation
for Quaternionic Quantum Mechanics.

\section{Commutation Relations and Time Evolution for Operators}

The canonical (anti)commutation relations can be obtained from the action
using the canonical infinitesimal generator,\footnote{%
We are adopting the sum convention.}%
\begin{equation*}
\hat{G}=\hat{\iota}\hat{p}_{r}\delta \hat{q}^{r}
\end{equation*}%
from which we extract the following set of functional relationships:%
\begin{align*}
\delta \hat{q}^{s}& =-\left[ \hat{q}^{s},\hat{\iota}\right] \hat{p}%
_{r}\delta \hat{q}^{r}-\hat{\iota}\left[ \hat{q}^{s},\hat{p}_{r}\right]
_{\mp }\delta \hat{q}^{r}\mp \hat{\iota}\hat{p}_{r}\left[ \hat{q}^{s},\delta
\hat{q}^{r}\right] _{\mp } \\
\hat{0}& =-\left[ \hat{p}_{s},\hat{\iota}\right] \hat{p}_{r}\delta \hat{q}%
^{r}-\hat{\iota}\left[ \hat{p}_{s},\hat{p}_{r}\right] _{\mp }\delta \hat{q}%
^{r}\mp \hat{\iota}\hat{p}_{r}\left[ \hat{p}_{s},\delta \hat{q}^{r}\right]
_{\mp } \\
\hat{0}& =\left[ \hat{q}^{s},\hat{\iota}\right] \delta \hat{p}_{r}\hat{q}%
^{r}+\hat{\iota}\left[ \hat{q}^{s},\delta \hat{p}_{r}\right] _{\mp }\hat{q}%
^{r}\pm \hat{\iota}\delta \hat{p}_{r}\left[ \hat{q}^{s},\hat{q}^{r}\right]
_{\mp } \\
\delta \hat{p}_{s}& =\left[ \hat{p}_{s},\hat{\iota}\right] \delta \hat{p}_{r}%
\hat{q}^{r}+\hat{\iota}\left[ \hat{p}_{s},\delta \hat{p}_{r}\right] _{\mp }%
\hat{q}^{r}\pm \hat{\iota}\delta \hat{p}_{r}\left[ \hat{p}_{s},\hat{q}^{r}%
\right] _{\mp }
\end{align*}%
This gives a system of equations between the canonical variables and their
variations whose formal solution is unknown. One possible solution is to
choose infinitesimal variations in order that
\begin{align*}
\left[ \hat{q}^{s},\delta \hat{q}^{r}\right] _{\mp }& =\left[ \hat{p}%
_{s},\delta \hat{q}^{r}\right] _{\mp }=\hat{0} \\
\left[ \hat{q}^{s},\delta \hat{p}_{r}\right] _{\mp }& =\left[ \hat{p}%
_{s},\delta \hat{p}_{r}\right] _{\mp }=\hat{0}
\end{align*}

However, terms involving the (anti)commutator of $\hat{\iota}$\ still
remain, which could imply in \textquotedblleft deviations\textquotedblright\
from the canonical commutation relations. That is why in \cite{FinkQ1} is
adopted the superselection rule%
\begin{equation}
\left[ \hat{q}^{s},\hat{\iota}\right] =\left[ \hat{p}_{s},\hat{\iota }\right]
=\hat{0}   \label{SuperSelcPQ}
\end{equation}
which conduct to%
\begin{align*}
\left[ \hat{p}_{s},\hat{p}_{r}\right] _{\mp} & =\hat{0} \\
\left[ \hat{q}^{s},\hat{q}^{r}\right] _{\mp} & =\hat{0} \\
-\hat{\iota}\left[ \hat{q}^{s},\hat{p}_{r}\right] _{\mp} & =\delta _{\,r}^{s}
\end{align*}

To obtain an expression closer to the complex case, let us to suppose that
the anti-hermitean operator $\hat{\iota}$\ is also \emph{unitary}. By this
way,%
\begin{align*}
\left[ \hat{p}_{s},\hat{p}_{r}\right] _{\mp} & =\hat{0} \\
\left[ \hat{q}^{s},\hat{q}^{r}\right] _{\mp} & =\hat{0} \\
\left[ \hat{q}^{s},\hat{p}_{r}\right] _{\mp} & =\hat{\iota}\delta_{\,r}^{s}
\end{align*}

This means that to obtain the standard form of the Heisenberg algebra for
the canonical variables $\hat{q}$\ and $\hat{p}$ one shall to demand \emph{%
both} conditions 2 and 3 from section \ref{Condicoes}.

The equation of motion for operators can also be obtained from the
variational principle doing variations only in the temporal parameter,%
\begin{equation*}
\frac{\vec{d}\hat{A}}{\vec{d}t}=\hat{\iota}\left[ \hat{A},\hat{H}\right] +%
\frac{\vec{\partial}\hat{A}}{\vec{\partial}t}
\end{equation*}

\subsection{Application: The Quaternionic Harmonic Oscillator}

Assuming that a quaternionic harmonic oscillator is described by the
following Lagrangian operator\footnote{%
Simplifying notation we will omit the symbol $\widehat{}$ from the operator
in this section. We maintain it only over the imaginary unity in order to
reinforce that here it is an \emph{operator}.}%
\begin{equation*}
L=\frac{1}{2}\left( \dot{q}^{\dagger}\dot{q}-\omega^{2}q^{\dagger}q\right)
,\quad q=\sum_{\alpha=0}^{3}q^{\alpha}e_{\alpha}
\end{equation*}

Taking functional variations of this operator, we find%
\begin{align*}
\delta L & =\frac{1}{2}\left( \left( \delta\dot{q}^{\dagger}\right) \dot{q}%
+\left( \dot{q}^{\dagger}\right) \delta\dot{q}-\omega^{2}\left[ \left(
\delta q^{\dagger}\right) q+q^{\dagger}\delta q\right] \right) = \\
& =\frac{1}{2}\left( \frac{d\left( \delta q^{\dagger}\dot{q}+\dot {q}%
^{\dagger}\delta q\right) }{dt}-\left[ \delta q^{\dagger}\left( \ddot {q}%
+\omega^{2}q\right) +\left( \ddot{q}^{\dagger}+\omega^{2}q^{\dagger }\right)
\delta q\right] \right)
\end{align*}

Therefore, the infinitesimal generator for the functional variations in the
fundamental operator is%
\begin{align*}
G & =\frac{1}{2}\hat{\iota}\left( \delta q^{\dagger}\dot{q}+\dot {q}%
^{\dagger}\delta q\right) \\
\bar{G} & =-\frac{1}{2}\hat{\iota}\left( q^{\dagger}\delta\dot{q}+\delta\dot{%
q}^{\dagger}q\right)
\end{align*}
whose induced variations are\footnote{%
The position of the indices is completely arbitrary here since we are
dealing with a cartesian space.}%
\begin{align*}
\frac{1}{2}\delta q^{\beta} & =\frac{1}{2}\left( \left[ q^{\beta},\hat{\iota}%
\delta q^{\alpha\dagger}\dot{q}_{\alpha}\right] +\left[ q^{\beta},\hat{\iota}%
\dot{q}_{\alpha}^{\dagger}\delta q^{\alpha}\right] \right) = \\
& =-\frac{1}{2}\left( \hat{\iota}\delta q^{\alpha\dagger}\left[ q^{\beta },%
\dot{q}_{\alpha}\right] +\hat{\iota}\left[ q^{\beta},\dot{q}_{\alpha
}^{\dagger}\right] \delta q^{\alpha}\right)
\end{align*}%
\begin{align*}
\delta q^{\beta} & =-\hat{\iota}\left( \delta q^{\alpha\dagger}\left[
q^{\beta},\dot{q}_{\alpha}\right] +\left[ q^{\beta},\dot{q}_{\alpha
}^{\dagger}\right] \delta q^{\alpha}\right) \\
\delta q^{\beta\dagger} & =-\hat{\iota}\left( \delta q^{\alpha\dagger }\left[
q^{\beta\dagger},\dot{q}_{\alpha}\right] +\left[ q^{\beta\dagger },\dot{q}%
_{\alpha}^{\dagger}\right] \delta q^{\alpha}\right) \\
\delta\dot{q}^{\beta} & =\hat{\iota}\left( \delta\dot{q}_{\alpha}\left[ \dot{%
q}^{\beta},q^{\alpha\dagger}\right] +\left[ \dot{q}^{\beta},q^{\alpha }%
\right] \delta\dot{q}_{\alpha}^{\dagger}\right) \\
\delta\dot{q}^{\beta\dagger} & =\hat{\iota}\left( \delta\dot{q}_{\alpha }%
\left[ \dot{q}^{\beta\dagger},q^{\alpha\dagger}\right] +\left[ \dot {q}%
^{\beta\dagger},q^{\alpha}\right] \delta\dot{q}_{\alpha}^{\dagger}\right)
\end{align*}

Assuming that the operators $q^{\beta}$, $\dot{q}^{\beta}$, $\dot{q}%
^{\beta\dagger}$ and $q^{\beta\dagger}$\ are kinematically independent, we
have the canonical commutation relations,%
\begin{align*}
\left[ q^{\beta\dagger},\dot{q}_{\alpha}^{\dagger}\right] & =\left[
q^{\beta},\dot{q}_{\alpha}\right] =0 \\
\left[ q^{\beta\dagger},\dot{q}_{\alpha}\right] & =\left[ q^{\beta},\dot{q}%
_{\alpha}^{\dagger}\right] =\hat{\iota}\delta_{\,\alpha}^{\beta}
\end{align*}

\section{Schr\"{o}dinger Equation and the Coordinate Representation\label%
{IConst}}

Taking variations only over the final state in a given transition,%
\begin{align*}
\delta\left\vert b_{t_{1}}\right\rangle & =0\rightarrow\delta\mathbf{\hat {q}%
}\left( t_{1}\right) =\mathbf{\hat{0}}\quad\delta t_{1}=0 \\
\delta\left\langle a_{t_{2}}\right\vert & \not =0\rightarrow\delta \mathbf{%
\hat{q}}\left( t_{2}\right) \not =\mathbf{\hat{0}}\quad\delta t_{2}\not =0
\end{align*}
we have
\begin{equation*}
\delta\left( \left\langle a_{t_{2}}|b_{t_{1}}\right\rangle \right)
=\left\langle a_{t_{2}}\right\vert \hat{\iota}\left( \mathbf{\hat{p}}%
_{2}\cdot\delta\mathbf{\hat{q}}_{2}-\hat{H}\delta t_{2}\right) \left\vert
b_{t_{1}}\right\rangle
\end{equation*}
Now, let us identify the description $a$ as the generalized coordinates,
i.e., the description where the operators $\mathbf{\hat{q}}$ are diagonal,
and the state $\left\vert b_{t_{1}}\right\rangle $\ as an arbitrary state $%
\left\vert \Psi\right\rangle $. From the commutation relations deduced
before we have
\begin{align*}
\delta\left( \left\langle q_{t_{2}}|\Psi\right\rangle \right) &
=\left\langle q_{t_{2}}\right\vert \delta\mathbf{\hat{q}}_{2}\cdot \mathbf{%
\hat{p}}_{2}\hat{\iota}\left\vert \Psi\right\rangle -\left\langle
q_{t_{2}}\right\vert \hat{\iota}\hat{H}\delta t_{2}\left\vert \Psi
\right\rangle = \\
& =\delta\mathbf{q}_{2}\cdot\left\langle q_{t_{2}}\right\vert \hat{\iota }%
\mathbf{\hat{p}}_{2}\left\vert \Psi\right\rangle -\delta t_{2}\left\langle
q_{t_{2}}\right\vert \hat{\iota}\hat{H}\left\vert \Psi\right\rangle
\end{align*}
But,
\begin{equation*}
\delta\left( \left\langle q_{t_{2}}|\Psi\right\rangle \right) =\delta
\mathbf{q}_{2}\cdot\frac{\partial\left\langle q_{t_{2}}|\Psi\right\rangle }{%
\partial\mathbf{q}_{2}}+\delta t_{2}\frac{\partial\left\langle
q_{t_{2}}|\Psi\right\rangle }{\partial t_{2}}
\end{equation*}
then,%
\begin{align*}
\frac{\partial\left\langle q_{t_{2}}|\Psi\right\rangle }{\partial \mathbf{q}%
_{2}} & =\left\langle q_{t_{2}}\right\vert \hat{\iota}\mathbf{\hat{p}}%
_{2}\left\vert \Psi\right\rangle \\
\frac{\partial\left\langle q_{t_{2}}|\Psi\right\rangle }{\partial t_{2}} & =%
\mathbf{-}\left\langle q_{t_{2}}\right\vert \hat{\iota}\hat{H}\left\vert
\Psi\right\rangle
\end{align*}

Inserting a completeness relation for the coordinate eigenstates, we find%
\begin{align*}
\frac{\partial\left\langle q_{t_{2}}|\Psi\right\rangle }{\partial \mathbf{q}%
_{2}} & =\int d\bar{q}\left\langle q_{t_{2}}\right\vert \hat{\iota }%
\left\vert \bar{q}_{t_{2}}\right\rangle \left\langle \bar{q}%
_{t_{2}}\right\vert \mathbf{\hat{p}}_{2}\left\vert \Psi\right\rangle \\
\frac{\partial\left\langle q_{t_{2}}|\Psi\right\rangle }{\partial t_{2}} & =%
\mathbf{-}\int d\bar{q}\left\langle q_{t_{2}}\right\vert \hat{\iota }%
\left\vert \bar{q}_{t_{2}}\right\rangle \left\langle \bar{q}%
_{t_{2}}\right\vert \hat{H}\left\vert \Psi\right\rangle
\end{align*}
The first equation\footnote{%
Note that we have made use of the fact that the spectrum of the coordinates
is \emph{real}.} gives the representation of the momentum operator in the
coordinate representation assuming that the spectrum of $\hat{\iota}$\ is
know, while the second is the\emph{\ Schr\"{o}dinger equation}.

If, by hypothesis, the operator $\hat{\iota}$\ has always the same value in
any point of the coordinate space and at any instant of time, then
\begin{subequations}
\label{EqSchQConst}
\begin{align}
\frac{\partial\left\langle q_{t_{2}}|\Psi\right\rangle }{\partial \mathbf{q}%
_{2}} & =\iota\left\langle q_{t_{2}}\right\vert \mathbf{\hat{p}}%
_{2}\left\vert \Psi\right\rangle  \label{EqSchQConst1} \\
\frac{\partial\left\langle q_{t_{2}}|\Psi\right\rangle }{\partial t_{2}} & =%
\mathbf{-}\iota\left\langle q_{t_{2}}\right\vert \hat{H}\left\vert
\Psi\right\rangle   \label{EqSchQConst2}
\end{align}
where $\iota$\ is the expected value of $\hat{\iota}$. Of course, this last
hypothesis is contained in the statement 2 of the section \ref{Condicoes}
and it imply that the \emph{operator }$\hat{\iota}$ is actually a constant
imaginary pure quaternion.

\section{Final Remarks}

The Schwinger Measurement Algebra formulation for quantum kinematics is a
powerful tool to disconnect the physical contents in quantum measurements
from the mathematical requirements of consistence. At same time, it provides
a natural way to achieve generalizations of standard Quantum Mechanics and
provide a clear view of the price paid for such generalizations.

In particular, besides we have found difficulties to construct a linear
functional relating operators and values in the quaternionic ring, it was
still possible to achieve a well defined statistical interpretation for
Quaternionic Quantum Mechanics. The essential elements for such construction
are the noncompatibility of successive measurements, providing the
fundamental law of multiplication for measurement symbols, and the
automorphism $\left\langle a|b\right\rangle
\rightarrow\lambda_{a}\left\langle a|b\right\rangle \lambda_{b}^{-1}$\ of
the scalar ring. In principle, any theory with these basic characteristics
can also have an statistical interpretation. Notwithstanding, for an
appropriate interpretation, some additional properties are required for the
probability measure $p\left( a|b\right) $, such as the conservation of their
associated current in a closed system. In fact, it is the essential feature
that Adler used to prove the non-extensivity of Quantum Mechanics for
octonions or higher dimension hypercomplex numbers \cite{Adler}.

It must be stressed that there are several problems which are not
investigated above, such as the effect of the superselection rules (\ref%
{SuperSelcPQ}) over representations of the canonical variables $\hat{p}$\
and $\hat{q}$, or the physical effects of the new quaternionic degrees of
freedom.

Although we have not treated composite systems (i.e., many particle systems)
it is possible to advance some characteristics which should originate from
the quaternionic noncommutativity. It is well known that in the classical
physics there are no phase relations to be considered among subsystems of a
bigger system (non-interacting particles) if we sum or multiply (by
Cartesian product) their phase spaces. In Complex Quantum Mechanics there
are phase relations between states which are important if we sum their state
spaces, but they are not important for the product of such spaces
(understood as a tensorial product). In Quaternionic Quantum Mechanics these
phase relations should be important whatever one is dealing with sum or
product of spaces, since the phase factor now is a quaternion.

This new feature of Quaternionic Quantum Mechanics can be interpreted in
terms of a complementarity argument. In classical physics does exist
complementarity relations because all measurements can, in principle, have
an infinity precision. In the real and complex quantum mechanics there are
complementarity relations among physical properties of the same system, but
not between properties of different non-interacting systems. In quaternionic
quantum mechanics there are complementarity between some properties for any
pair of physical systems or subsystems. This is because the phase factor $%
e^{\varphi\left( a\right) }$\ can not be additively composed when
multiplying quaternions, as specified in the section \ref{estatistica}.
Therefore, there is no reasonable way to form composite systems in order to
have all observables associated in one system to be compatible with all the
observables in any other systems or, in other words, to commute with all the
others observables in different systems. Actually, one can expect this to be
greater difficulty to describe many particle systems in quaternionic quantum
theory

It is important to observe that besides the notion of a quaternionic Hilbert
space has been a little vague here it is possible to develop the concepts
the Geometry of States, as done by Schwinger \cite{Schwinger}, for the
quaternionic ring. The idea and properties of such vectorial space emerge
naturally in the Geometry of States. This was not done here simply by
matters of space and convenience since that we were interested not only on
the kinematical side, but also in the dynamic aspects of the quaternionic
theory. For those interested in the spectral theory of quaternionic Hilbert
spaces is interesting to check \cite{FinkQ2} where the main theorems and
ideas are introduced with a pedagogical explanation of how to perform the
calculations in a vectorial space of scalars in $\mathbb{H}$.

With respect to quartenionic quantum mechanics of a single particle one can
observe that the points where the operator $\hat{\iota}$\ appears are
essentially the same where the Planck constant $\hbar$ should be. Of course,
using a different system of units, one realizes that the operator $\hat{%
\iota }$\ takes the place of the combination $i/\hbar$ accordingly the
analogy applied here. By this way, the introduction of operators which fail
to commute with $\hat{\iota}$ can be understood as to promote the Planck
\textquotedblleft constant\textquotedblright\ to a new dynamic variable,
being interesting to investigate the fluctuations in the quantum of action
in such case. On the other hand, the superselection rule expressed by the
second condition in the section \ref{Condicoes} together with the hypothesis
made in the final of the section \ref{IConst} gave a \emph{classical}
meaning to $\hat{\iota}$ excluding the interference between their different
states. This is equivalent to \textquotedblleft freeze\textquotedblright\
the actual value of the imaginary unity operator suppressing this new
possibilities. Therefore, we find a natural extension of the equations (\ref%
{EqSchQConst}) admitting that the operator $\hat{\iota}$\ actually is a new
\emph{fundamental field}, i.e., a new dynamic variable which depends from
the space-time point where it is observed. This idea was partially developed
in \cite{FinkQ3} where it is proposed a quaternionic general covariance
principle, which means a theory for the parallel transport of quaternions
over a manifold, and a field equation for the operator $\hat{\iota}$. One of
the most surprising results of this theory is that the field equations
obtained are very similar to the electromagnetic ones but with \emph{three}
fundamental vectorial bosons, one neutral and massless and two others
massive and charged. So, Quaternionic Quantum Mechanics could be considered
one of the first attempts to construct an unified theory for the electroweak
interactions (1963) and perhaps could model at least a sector of the
complete electroweak interactions.

\subsection*{Acknowledgments}

The authors are grateful to our friend Prof. Rold\~{a}o da Rocha for useful
discussions. BMP also thanks CNPq-Brazil for partial support.
\end{subequations}

\end{document}